\documentclass[12pt]{article}
\usepackage{ametsoc,lineno,bbm}
\usepackage{xspace}
\usepackage{url,color,rotating}

\usepackage[utf8]{inputenc}

\newcommand{\Z}{\mathbb{Z}}

\newcommand{\cov}{\mbox{cov}}

\newcommand{\myabstract}{
{The purpose of this study is to perform verification of the structural characteristics of high-resolution spatial forecasts without relying on an object identification algorithm. To this end, 
 a wavelet approach developed for image texture analysis is applied to an ensemble of high-resolution quantitative precipitation forecasts. 
The forecasts are verified against estimates from a high-resolution regional reanalysis  with a similar model version. The wavelet approach estimates an averaged wavelet spectrum for each spatial field of the ensemble forecasts and the reanalysis, thereby removing all information on the localization of precipitation and investigating solely the overall structure of forecasts and reanalysis. {In order to assess skill using a multivariate score, an additional reduction of dimensionality is needed. This is performed using singular vectors from a linear discriminant analysis as it favors data compression in the direction where the ensemble is most discriminating. } We discuss implications of this strategy, show that  the averaged wavelet spectra give valuable information on forecast performance. 
The skill difference between a so-called perfect forecast using for verification a member of the ensemble, and the non-perfect forecast using the reanalysis points to significant deficiencies of the forecast ensemble. {Overall, the discriminating power solely based on global spectral information is remarkable, and the COSMO-DE-EPS is a quite good forecast ensemble with respect to the reanalysis.}%The discriminating power for the reanalysis solely on the basis of the averaged wavelet spectra is about 80\%, and is as high as for the ensemble itself.
}
}

\begin{document}

%%%%%%%%%%%%%%%%%%%%%%%%%%%%%%%%%%%%%%%%%%%%%%%%%%%%%%%%%%%%%%%%%%%%%
% TITLE
%
% Enter your TITLE here
%%%%%%%%%%%%%%%%%%%%%%%%%%%%%%%%%%%%%%%%%%%%%%%%%%%%%%%%%%%%%%%%%%%%%
\title{\textbf{\large{Spatial verification of high-resolution ensemble precipitation forecasts using local wavelet spectra}}}
%
% Author names, with corresponding author information. 
% [Update and move the \thanks{...} block as appropriate.]
%
\author{\textsc{Florian Kapp, Petra Friederichs}\thanks{\textit{Corresponding author address:} 
				Petra Friederichs, Meteorological Institute, University of Bonn, 
				Auf dem H\"ugel 20, 53121 Bonn, Germany. 
				\newline{E-mail: pfried@uni-bonn.de}}
%				\quad
\textsc{, Sebastian Brune} \\ \textsc{ Michael Weniger}
\\
\textit{\footnotesize{Meteorological Institute, University of Bonn,  Germany}}
}
%
% The following block of code will handle the formatting of the title page depnding on whether
% we are formatting a double column (dc) author draft or a single column paper for submission.
% AUTHORS SHOULD SKIP OVER THIS... There is nothing to do in this section of code.
\ifthenelse{\boolean{dc}}
{
\twocolumn[
\begin{@twocolumnfalse}
\amstitle

% Start Abstract (Enter your Abstract above.  Do not enter any text here)
\begin{center}
\begin{minipage}{13.0cm}
\begin{abstract}
	\myabstract
	\newline
	\begin{center}
		\rule{38mm}{0.2mm}
	\end{center}
\end{abstract}
\end{minipage}
\end{center}
\end{@twocolumnfalse}
]
}
{
\amstitle
\begin{abstract}
\myabstract

\end{abstract}
\newpage
}

%%%%%%%%%%%%%%%%%%%%%%%%%%%%%%%%%%%%%%%%%%%%%%%%%%%%%%%%%%%%%%%%%%%%%
% MAIN BODY OF PAPER
%%%%%%%%%%%%%%%%%%%%%%%%%%%%%%%%%%%%%%%%%%%%%%%%%%%%%%%%%%%%%%%%%%%%%

\section{Introduction}

Convective precipitation within thunderstorms or showers is a very local mesoscale phenomenon. Quantitative prediction of convective precipitation is still very challenging. High-resolution limited area numerical weather prediction models (NWP) aim at the prediction of such convective events and their impacts \citep[cf.][]{stensrud2009}. However, high-resolution forecasts are not able to precisely locate showers and storms in space and/or time.  {Calculating skill measures such as the mean square error (MSE) between two fields results in a {double penalty} \citep{Ebert2008}, where the penalization results from both missing the event in one place and falsely predicting it in another.} 
 Thus, despite their ability to simulate small scale features, gridpoint-by-gridpoint verification methods suggest that higher resolved models generate worse forecasts \citep{gilleland2009}.  Furthermore, these scores suffer from a lack of important diagnostic information about the type of errors such as displacement and structural differences of features, {and do not  provide information on the scale at which a forecast is skillful.} 

{Different verification techniques have been developed to address these problems. In the overview by \citet{gilleland2009} several spatial verification methods are compared as part of a coordinated intercomparison project. Four classes of spatial verification methods were identified: neighborhood, feature-based, field deformation and scale-separation.}
  Neighborhood methods take adjacent gridpoints into account, for example by upscaling the fields, thereby attenuating the double penalty problem  \citep{theis2005,Ebert2009}. However, such a fuzzy method smears out small-scale variability and {reduces accuracy of the forecasts.} 
 The fraction skill score {relies on} such a neighborhood verification technique \citep[e.g.][]{roberts2008}.
Feature based approaches search the fields for objects such as single storms and compare the characteristics of the objects \citep[e.g.][]{marzban2006,davis2006}. \citet{Ebert2000} for example determine contiguous grid points at which a threshold is exceeded, so that information about size, orientation and location amongst others can be retrieved. The so-called SAL score of \citet{wernli2008} investigates the objects' structure, amplitude and location, and has found application in the evaluation of, e.g., cloud-precipitation features in large-eddy simulations \citep{heinze2017}. {\citet{weniger2016} show that feature based approaches may become very sensitive to the {parameter choice of the threshold-based} object identification method}, and may provide very different scores particularly in presence of large objects. {In this respect, they investigate the discriminating
power of SAL and {show} that changes in the parameters may have
larger effects on the object-dependent SAL scores than  a complete loss of
temporal collocation.}
Another interesting method for forecast verification is related to image warping. This method deforms the two fields/objects such that some distance measure is minimized. It may be applied to single features \citep{alexander1999} or fields {\citep{gilleland2010}}. 

{Scale-separation techniques use single-band spatial filters, e.g. via spherical harmonics, wavelets or Fourier transforms, to assess forecast performance at different scales \citep[e.g.][]{Jung2008}. Scale-separation techniques that completely ignore any information on the location are based on the assessment of the spectra  \citep[e.g.][]{harris2001}  or on the estimate of the spatial covariance function \citep[e.g.][]{marzban2009,scheuerer2015}.}  {\citet[][]{lack2010} apply Fourier transform as part of a multi-scale object identification scheme. \citet{Keil2009} apply a spatial filter prior to performing a field-morphing verification. These approaches are combinations of a scale-separation method with a feature based and field-deformation technique, respectively.}

Wavelets have proven to be a very effective tool in image processing. In the context of spatial verification they offer a number of appealing characteristics. The ability to {reduce the dimensionality of data via scale-separation} is certainly one of the principal reasons for the success of wavelet transforms. Moreover, wavelets are localized in space and therefore, contrary to Fourier based approaches, do not require stationarity of the data.  A prime example for wavelet based verification is the intensity-scale skill score \citep[ISS,][]{casati2004,casati2007,casati2010}, which is based on the work of \citet{briggs1997} and uses a two-dimensional discrete wavelet transform for an orthogonal scale decomposition and defines a skill score based on the MSE. A comprehensive review of wavelet based methods for spatial verification is conducted in \citet{weniger2017}.

 {Most spatial verification methods are applied locally thereby ignoring the multivariate character of the data. For a multivariate verification, a drastic reduction of spatial degrees of freedom (DOF) is required prior to multivariate score estimation. The data reduction approach may be guided by an interest in special aspects. In our case we aim at assessing the spatial covariance structure of the forecasts compared to observations. This is what the texture analysis of \citet{eckley2010} is developed for.}
{The basic idea of \citet{eckley2010} is to use estimates of the local wavelet spectrum in order to compare images of different texture.} {They use a discrete two-dimensional wavelet transform that provides wavelet coefficients} of three directional {single-band} pass filters (horizontal, vertical, diagonal) on the whole field for each scale. 
We follow up on \citet{weniger2017} who provide a first attempt to use  Eckley's  texture analysis for spatial forecast verification {of precipitation and explore its potential for skill assessment, where skill solely relies on the  covariance structure of the spatial fields. We thereby intentionally ignore skill that may be assessed via standard methods such as MSE.}

Eckley's texture analysis is applied to an ensemble of forecasts with the high-resolution NWP  model COSMO-DE\footnote{Consortium of Small-scale Modeling, http://www.cosmo-model.org/.}, operated by the German Meteorological Service \citep[COSMO-DE-EPS,][]{baldauf2011,gebhardt2011,peralta2012}. {As observations we use} the regional reanalysis COSMO-REA2 \citep{wahl2017}, where latent heat nudging is used to assimilate precipitation estimates from radar observations.
Our investigation concentrates on 14 days in 2011 with considerable precipitation over Germany due to convective, stratiform and mixed convective-stratiform precipitation. 
{We assess the following aspects. First of all, we provide a multivariate forecast distribution in a subspace of the averaged wavelet spectra. The subspace is defined using a linear discriminant analysis (LDA), where the  ensemble members for one day build a group, and the new data is the respective reanalysis valid at the same time as the forecast. 
The LDA subspace optimally discriminates the 14 forecast days in the {ensemble data}. The forecast  skill of the ensemble is assessed within this subspace using the logarithmic score. Finally, new data  is attributed to a group (i.e.~the ensemble forecast) on the basis of the posterior probability of the group of forecasts given the new data. With this procedure we assess whether with the LDA subspace of the averaged wavelet spectra the ensemble provides a good forecast and is discriminating in the sense of \citet{Murphy1987}  with respect to the reanalysis.}

The remainder of this article is structured as follows.
We first introduce in Section \ref{Sec:Data} the ensemble forecast and reanalysis data used in this study.
Section \ref{Sec:Method} provides a short introduction to the texture analysis of \citet{eckley2010} and discusses our verification strategy. A more detailed mathematical review of Eckley's texture analysis is presented in the appendix A.
In section \ref{Sec:Results} we compare the skill of a perfect forecast and the reanalysis and investigate the attribution ability on the basis of the averaged wavelet spectra. Section \ref{Sec:Conclusions} concludes the study and discusses possible extensions of this approach.

\section{Data}\label{Sec:Data}

Ensemble forecasts of hourly precipitation are provided by the high-resolution COSMO-DE ensemble prediction system  \citep[COSMO-DE-EPS,][]{baldauf2011,gebhardt2011,peralta2012}.  The 20 member ensemble is operated by the German Meteorological Service (Deutscher Wetterdienst, DWD) since May 2012 covering Germany and neighboring countries with a grid of approximately 2.8 km horizontal distance and 50 vertical levels. 
{The COSMO-DE-EPS is set up using four global forecasts from different meteorological services. Each of the four global forecasts forces five ensemble members with perturbed parameterizations and initial perturbations. For more details see \citet{peralta2012}.}

As observational data we take the COSMO-REA2 reanalysis \citep{wahl2017} in order to avoid missing data. COSMO-REA2 and COSMO-DE-EPS share a similar NWP model, where COSMO-REA2 operates on a 2 km and COSMO-DE on a 2.8 km grid.  COSMO-REA2  hourly precipitation totals are interpolated on COSMO-DE 2.8 km grid using the nearest neighbor method. The reanalysis uses latent heat nudging to assimilate radar precipitation estimates. We thus expect the reanalysis to be very close to observations in this respect. 

We {selected} 14 forecasts for precipitation days in the period from June to October 2011 (Tab. \ref{casestab}), all initialized at 00 UTC. For each day, a late afternoon hour is chosen. The cases provide a range of precipitation fields ranging from weakly forced and highly scattered convective days such as 5 June 2011 to frontal cases such as 22 June 2011.  {These two specific cases} are intensively studied in \citet{weijenborg2015} and \citet{weijenborg2017}.

\section{Methods}\label{Sec:Method}
\subsection{Motivation}

{Verification of high-dimensional spatial fields (i.e. the comparison  high-dimensional multivariate random fields) is only meaningful in a reduced space with a small number of DOF. Otherwise, interpretation is very difficult as inference normally fails a multivariate significance test \citep{Hasselmann1979}. Further, measures such as the Mahalanobis distance (i.e. the log-likelihood under the assumption of a  multivariate Gaussian distribution) require the estimation of the inverse covariance matrix, which in turn is only possible in a space of a dimension that is smaller than the sample size or using restrictive distributional assumptions.} {Thus, forecast verification in high-dimensional spaces is above all a data compression problem.}

{Data compression may be motivated by looking at special characteristics of interest. In our case, we are searching for an alternative to the structure score of SAL, which aims at assessing structural differences between forecasts and observations. This is exactly what the texture analysis of \citet{eckley2010} is developed for.}

\subsection{Spectral analysis using discrete wavelet transforms}\label{Sec:Wavelets}

{We now provide a concise description of the general framework of the texture analysis as developed in \citet{eckley2010} and \citet{eckley2011}. A more detailed {version} is provided in the Appendix A.} 
{In a nutshell,} \citet{eckley2010} introduce a locally stationary two-dimensional stochastic wavelet process (LS2W, {Eq.~\ref{A6})}. {Local stationarity means that, although the LS2W process is non-stationary in space, the limit process when zooming into space at one location  becomes stationary {(Eq.~\ref{A7})}. Local stationarity corresponds to the assumption, that the spatial covariance structure of the process only varies slowly in space.  This process is assumed to produce the observations: in our application the LS2W process represents the precipitation process. In general meteorological data have a spatially non-stationary covariance structure, but that the spatial covariance structure of the process varies slowly in space}
 is a reasonable assumption for meteorological processes. 

The mathematical definition of a LS2W {(Eq.~\ref{A9})} uses {a set of orthonormal random variables, non-decimated two-dimensional wavelets {(introduced in the appendix A, Eq.~\ref{A3})} and wavelet coefficients} at each scale, direction and location, respectively. {\citet{eckley2010} use the formulation of the LS2W for the mathematical derivation of an estimator of the local autocovariance function and  its spectral decomposition.}
There are various methods to estimate the locally stationary spatial covariance function of a spatial process, e.g. by pooling the data around a location.  
The advantage of the {wavelet approach} of \citet{eckley2010} is that {the projection into the wavelet space and the estimation of the local wavelet spectrum corresponds to} a scale dependent pooling in space.

The estimation procedure in \citet{eckley2010} is as follows. 
A local wavelet periodogram (LWP) is defined as the squared values of the non-decimated two-dimensional wavelet coefficients {$I_{j,\mathbf r}^l$ (Eq.~\ref{A10})}. However, the LWP is an inconsistent and biased estimator of the local wavelet spectrum. Thus \citet{eckley2010}  use a smoothed LWP to obtain a consistent estimator, and then remove the bias using an operator matrix {$\mathbf A$ (Eq.~\ref{A9})}, which depends on the correlation between the wavelets at different  scales and directions. {The bias correction of the local wavelet spectra may result in negative spectral energy values}. Details on the approach are given in appendix A {and references therein}.

\subsection{Wavelet transformation of precipitation fields}\label{Sec:Application}
{For the application of wavelet transforms a few further preparations and decisions are to be made.} The two-dimensional discrete wavelet decomposition requires periodic boundary conditions. Thus,  all fields are padded with {about 250} zeros on each side to obtain fields with grid size of a power of 2, namely $2^{10} \times 2^{10}$ grid points. 
Beforehand, the  forecasts and observations {are smoothed at the boundaries} by a {linearly decreasing filter over 25 grid points ,} so that gradients between the precipitation field and the padding region flatten, which would otherwise induce artificial variance.

A non-decimated discrete wavelet transform is applied to each precipitation field independently, resulting in wavelet coefficients at each grid point for scales $j = 1, \ldots, 10$ and direction $l \in \{h,v,d \}$, respectively. We choose the Haar wavelet \citep{haar1910} since it makes interpretation of the decomposed fields easy. {A Haar wavelet decomposition is expected to capture the structure of the individual precipitation objects, as it is compatible with sharp boundaries.} The local wavelet spectra are estimated as described in Section \ref{Sec:Method} using the algorithm of \citet{eckley2010} and the R package \emph{LS2W}.

\begin{figure}
\centering
\includegraphics[height=0.9\textheight]{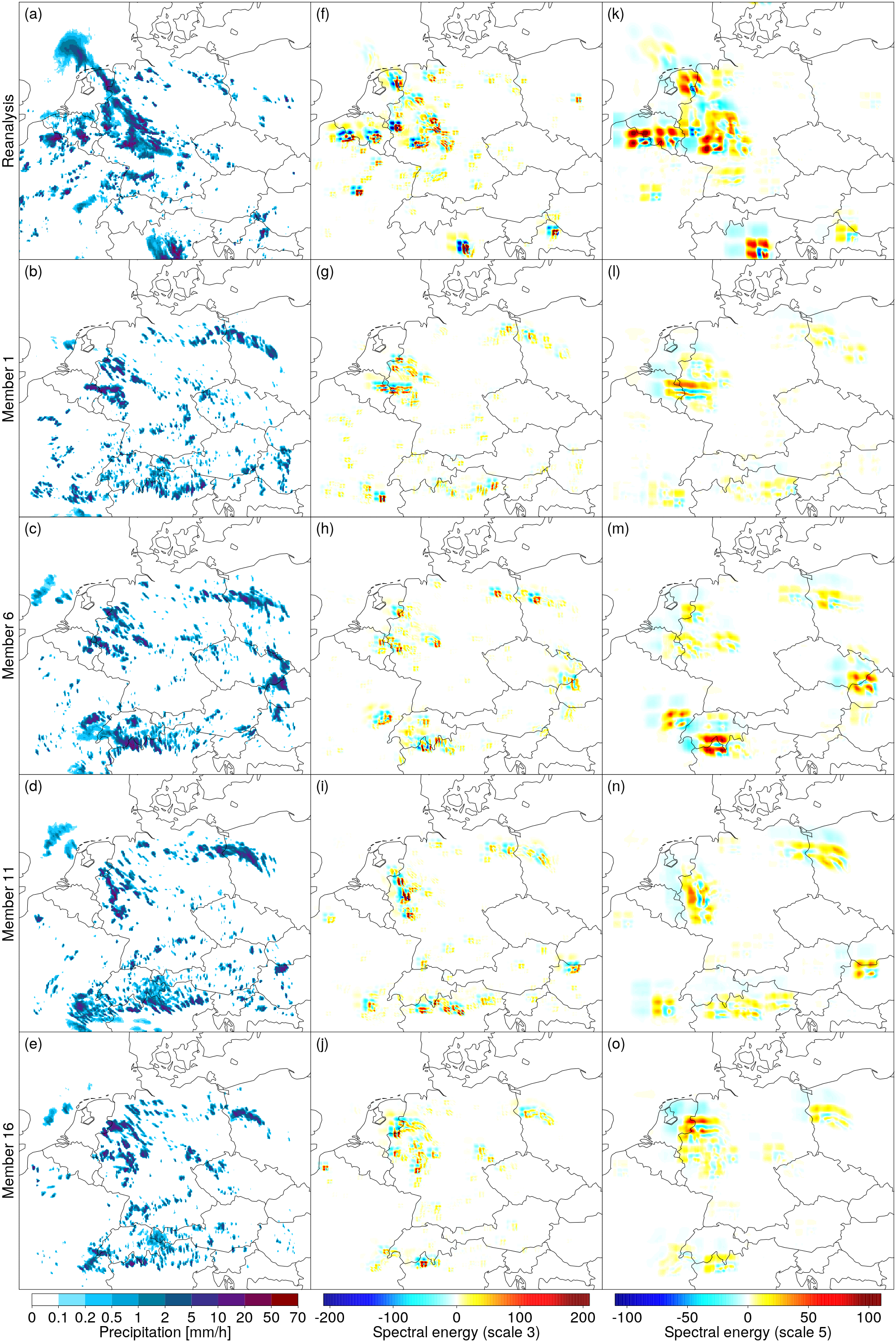}
 \caption{Precipitation and its local wavelet spectrum on 5 June 2011. (a)-(e) show precipitation fields in mm/h for (a) the reanalysis and (b)-(e) ensemble members 1, 6, 11, and 16, respectively. (f)-(j) show the respective unbiased wavelet spectra on scale 3 and (k)-(o) the respective unbiased wavelet spectra on scale 5 all averaged over the three directions.}
\label{overviewjun05}
\end{figure}

\begin{figure}
\centering
\includegraphics[height=0.9\textheight]{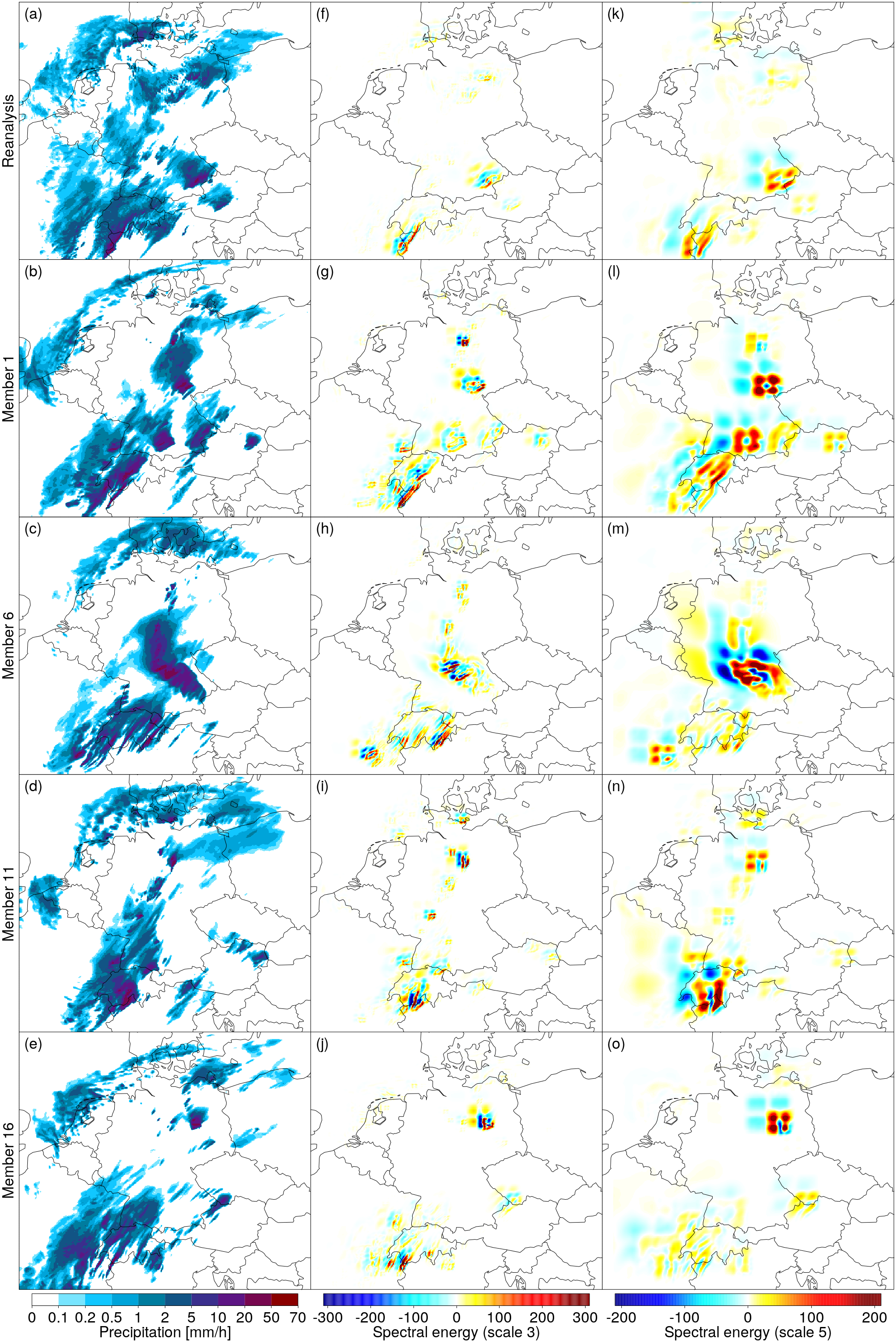}
 \caption{Same as Fig. \ref{overviewjun05} but for 13 July 2011.}
\label{overviewjul13}
\end{figure}

{For illustration Figs.~\ref{overviewjun05} and \ref{overviewjul13}  represent the precipitation fields together with the local  spectral energy  averaged over the three directions for scales $j=3$ and $j=5$ {, the dominant scales of precipitation of the respective day.}
On 5 June 2011 all fields display very scattered precipitation. While the ensemble members show very similar structures, the reanalysis shows more structure with enhanced precipitation over the western part of the domain. This is reflected in the spectral energy on both scales. Despite these differences, the overall structures are not very different.  On 13 July 2011 the structural difference between the ensemble simulations is larger, and in this respect, the reanalysis seems quite similar to the ensemble simulations.
Spectral energy, and hence total precipitation is larger on 13 July 2011 compared to 5 June 2011.
Note that the negative spectral energy is a consequence of the removal of the bias, and should not be interpreted.}

\begin{figure}
 \includegraphics[width=\textwidth]{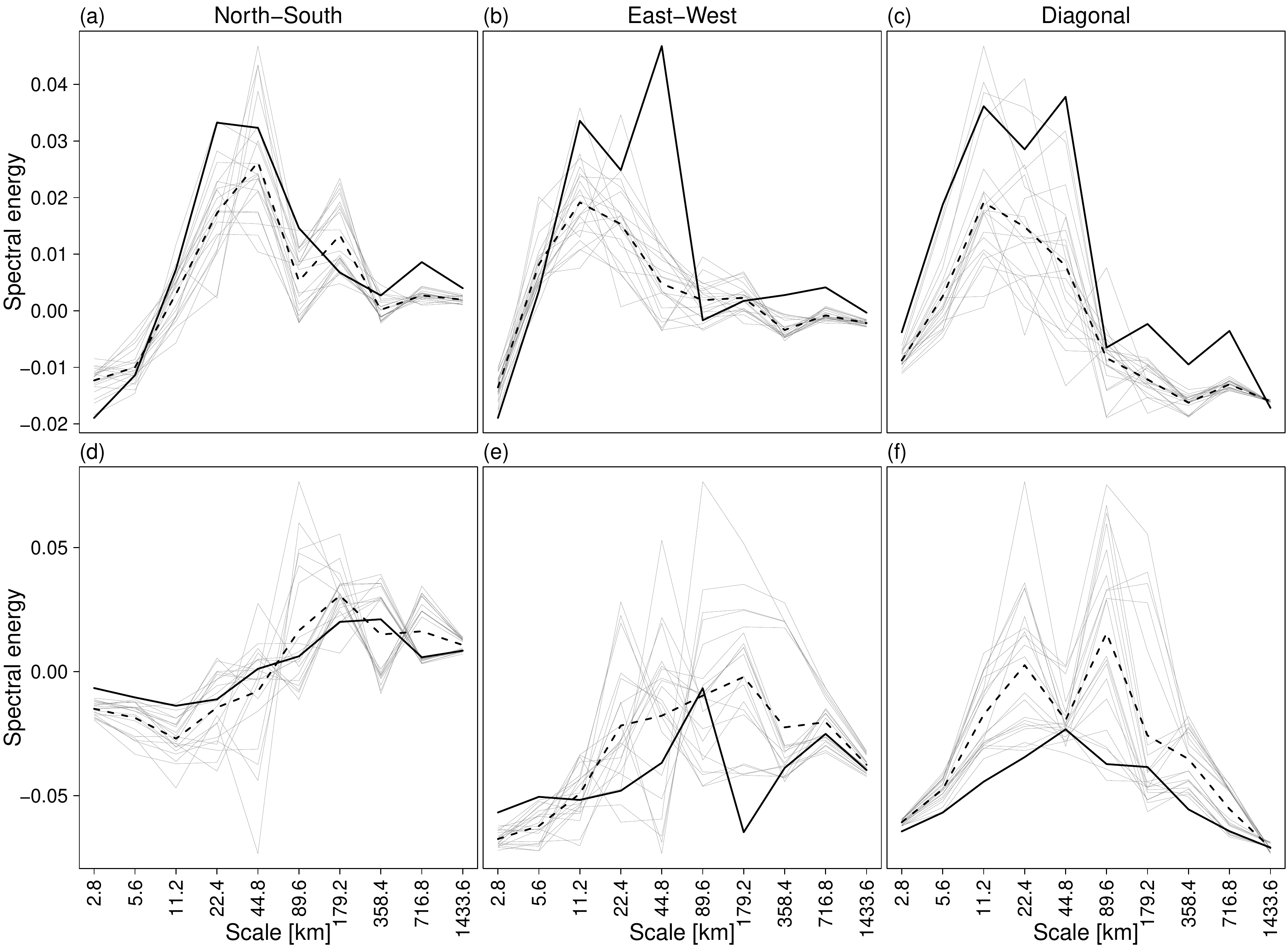}
 \caption{Spatially averaged wavelet spectra for (a)-(c) 5 June 2011 and (d)-(f) 13 July 2011 for the ensemble members (gray lines), and the reanalysis (solid black line). The dashed line indicates the average over the  wavelet spectra of the ensemble members.}
\label{Fig:Spectra}
\end{figure}

 {A wavelet decomposition does not result in a reduction of spatial DOF. The first drastic reduction is obtained by averaging the local wavelet spectrum over space. This results in a global wavelet spectrum with spectral energy for each scale (scales 1-10) and direction (North-South, East-West, and diagonal). For illustration the spatially averaged wavelet spectra of 5 June 2011 and 13 July 2011 are displayed in Fig. \ref{Fig:Spectra}. On 5 June 2011 maximum spectral energy is concentrated on scales 4 {(22.4 km)} to 5 {(44.8 km)}. The East-West and diagonal directions also show increase energy on scale 3 {(11.2 km)}, which is visible in Fig. \ref{overviewjun05} and probably a result of temporal averaging of {northward moving precipitation centers, stretching the precipitation fields more the North-South than in East-West direction}. The spectral energy of the reanalysis mostly lies within the spread of the ensemble, with one exception on East-West scale 5, where the reanalysis shows significantly more energy located over Western Germany and the Benelux states.
On 13 July 2011 spectral energy is maximal on scale 4 and larger {(22.4 km and above)}. Particularly scale 4 is most pronounced in diagonal direction. The spread of the ensemble is larger than on 5 June 2011, and the reanalysis mostly lies within the spread of the ensemble. The averaged wavelet spectra of 5 June 2011 and 13 July 2011 are clearly distinct in their characteristics.
 }
 The two largest scales, $j=9,10$, are largely influenced by the padding and are therefore discarded in the following analysis. {We also remove scales below 11 km (i.e.~scales $j=1,2$), as these scales lie below the effective resolution of the COSMO-DE-EPS model \citep{Bierdel2012}, and show only small spectral energy in Fig. \ref{Fig:Spectra}.}

\subsection{Linear discriminant analysis -- discrimination, attribution and scores}\label{Sec:MethodsLDA}
{The number of DOF of the averaged wavelet spectrum with 6 scales $\times$ 3 directions is still large compared to the sample size. We further reduce the number of DOF by using Fisher's LDA \citep[see textbook literature, e.g.,][for a comprehensive description of the method]{James2013}.  LDA searches for linear combinations (i.e. LDA vectors) that best discriminate among different classes $C_i$. In our case each day of the $N_c=14$ days constitutes a class $C_i, i=1,\ldots,N_c$, with $N_e=20$ ensemble forecasts $\mathbf m^{(j)}_i, j=1, \ldots,N_e$, and one observation $\mathbf o_i$, respectively. LDA  optimizes the ratio of between-class variance $S_B$ to within-class variance $S_W$, where $S_W$ is assumed to be equal for each class. This procedure is equivalent to an optimization of the posterior probability $p(C_i|\mathbf m^{(j)}_i)$ under the assumption of multivariate Gaussianity. Thus, LDA data compression uses the space spanned by vectors that maximize discrimination as a characteristic of ensemble forecasts only. The Gaussian assumption is quite reasonable for the averaged wavelet spectra, since already the estimation of the wavelet coefficients includes averaging, and then we average the wavelet spectra over a large number of grid points. Thus the central limit theorem justifies this assumption. The assumption of a constant within-class covariance matrix is necessary, since the number of predictors $N_e$ within each class is smaller than their size with 6 scales times 3 directions. A quadratic discriminant analysis (QDA) is more flexible than LDA in terms of discrimination, but only feasible within a smaller predictor space.}

{Given an observation $\mathbf o$, the LDA prediction is the posterior probability $p(C_i|\mathbf o)$, and should ideally be maximal if $\mathbf o \in C_i$. If this is the case, we say that cluster $C_i$ can be attributed to observation $\mathbf o$. The posterior probability $p(C_i|\mathbf o)$ is part of the discrimination term in Murphy's essay on ``What is a good forecast?'' \citep{Murphy1993}. The likelihood of an observation $\mathbf o$ to belong to class $C_i$ is $p(\mathbf o|C_i)$, where $p(\mathbf o|C_i)$ is a Gaussian with covariance  $S_W$ and mean $\sum_j \mathbf m^{(j)}_i/N_e$.  Since both measures $p(C_i|\mathbf o_i)$ and $p(\mathbf o|C_i)$ are related, an assessment of $p(C_i|\mathbf o_i)$ (i.e. how successful is attribution on the basis of $p(C_i|\mathbf o_i)$) provides information of the skill of a forecast system in the LDA subspace.}

{Our prediction approach uses cross-validation as follows. We randomly  select one ensemble member $\mathbf m^{(j_b)}_i$ of each day (i.e.~$j_b \in \{1, \ldots, N_c \}$ is random for each class $C_i$ and each cross-validation sample), respectively, where $b$ is the index of the cross-validation sample. The selected members are later used to assess the likelihood and do not enter the LDA. The LDA is thus applied to $N_c=14$ classes with each $N_e-1 = 19$ ensemble forecasts (i.e.~members $\mathbf m^{(j\neq j_b)}_i$). This provides us with 13 LDA vectors, which define a hierarchy of subspaces spanned by 1 to 13 LDA vectors. 
Given now the class average and the within-class covariance matrix in the respective LDA vector space, we calculate $\log p_b( \mathbf m^{(j_b)}_i|C_i)$, as well as $\log p_b( \mathbf m^{(j_b)}_k|C_i), \ k \neq i$. {The former} is the log-likelihood of a perfect forecast, since it is part of the ensemble, {the latter} of a random forecast which shares the statistical characteristics of the ensemble, but resolution is lost due to temporal re-sampling. 
Finally we calculate $\log p_b( \mathbf o_i|C_i)$, which is the log-likelihood of the reanalysis.
This procedure is repeated for $b=1, \ldots, N_b$ with $N_b=50$ to account for uncertainty with respect to the LDA vector estimation. 
}

{The sum over the likelihoods in the space spanned by a hierarchy of LDA vectors then constitutes a scoring rule of a probabilistic forecast in the respective LDA space. The logarithmic score is a proper scoring rule \citep{Gneiting2005}.  As reference score $S_{ref}$ we use the averaged log-likelihood of the ensemble members if the classes are different
\begin{equation}\label{Eq:Sref}
S_{ref} = \frac{1}{N_c (N_c-1) N_b} \sum^{N_c}_{i=1} \sum_{{k\neq i}}    \sum_{b=1}^{N_b} \log p_b( \mathbf m^{(j_b)}_k|C_i),
\end{equation}
the score of a perfect forecast $S_{perf}$ is defined as the averaged log-likelihood of the ensemble members if the classes are identical
  \begin{equation}
S_{perf,b} =   \frac{1}{N_c}\sum_{i=1}^{N_c}  \log p_b( \mathbf m^{(j_b)}_i|C_i),
\end{equation}
and the averaged log-likelihood of the reanalysis is given as
\begin{equation}
S_{obs} = \frac{1}{N_c N_b} \sum_{i=1}^{N_c} \sum_{b=1}^{N_b}  \log p_b( \mathbf o_i|C_i).
\end{equation}
The subscript $b$ denotes the $N_b=50$ cross-validation samples (see above).
If the reanalysis is indistinguishable from the ensemble forecasts, then the forecasts would be perfect, and skill is solely defined by the discrimination ability of the ensemble.
}

{We further assess the discrimination {probability} of correctly attributing new data {$\mathbf o_i$ ($\mathbf m_i^{(j_b)}$)} to the correct class $C_i$ as $\sum_i p_b(C_i|\mathbf o_i)/N_c$ ($\sum_i p_b(C_i|\mathbf m_i)/N_c$).
In order to avoid overfitting, we repeated the cross-validation procedure, but now we withhold the respective class $C_i$ when defining the LDA vectors and calculating $p_b(C_i|\mathbf o_i)$ or $p_b(C_i|\mathbf m_i)$.}

\section{Verification of averaged wavelet spectra}\label{Sec:Results}

Spatial averaging of the local wavelet spectra removes all information on the localization of the spectral energy, and solely investigates whether the overall structure at the different scales of forecasts and observations agrees. 
{We further  standardize the averaged wavelet spectra, i.e. subtract the averaged spectral energy over all scales and directions, and divide by the standard deviation over all scales and directions, thereby preserving the structural differences in the spectra. The reason is that COSMO-DE-EPS fails in the prediction of the total precipitation variance, which is equivalent to the total sum of the spectral energy (not shown). 
Why COSMO-REA2 and COSMO-DE-EPS are different in this respect is not clear.}

\subsection{Likelihood assessment}

\begin{figure}
 \includegraphics[width=\textwidth]{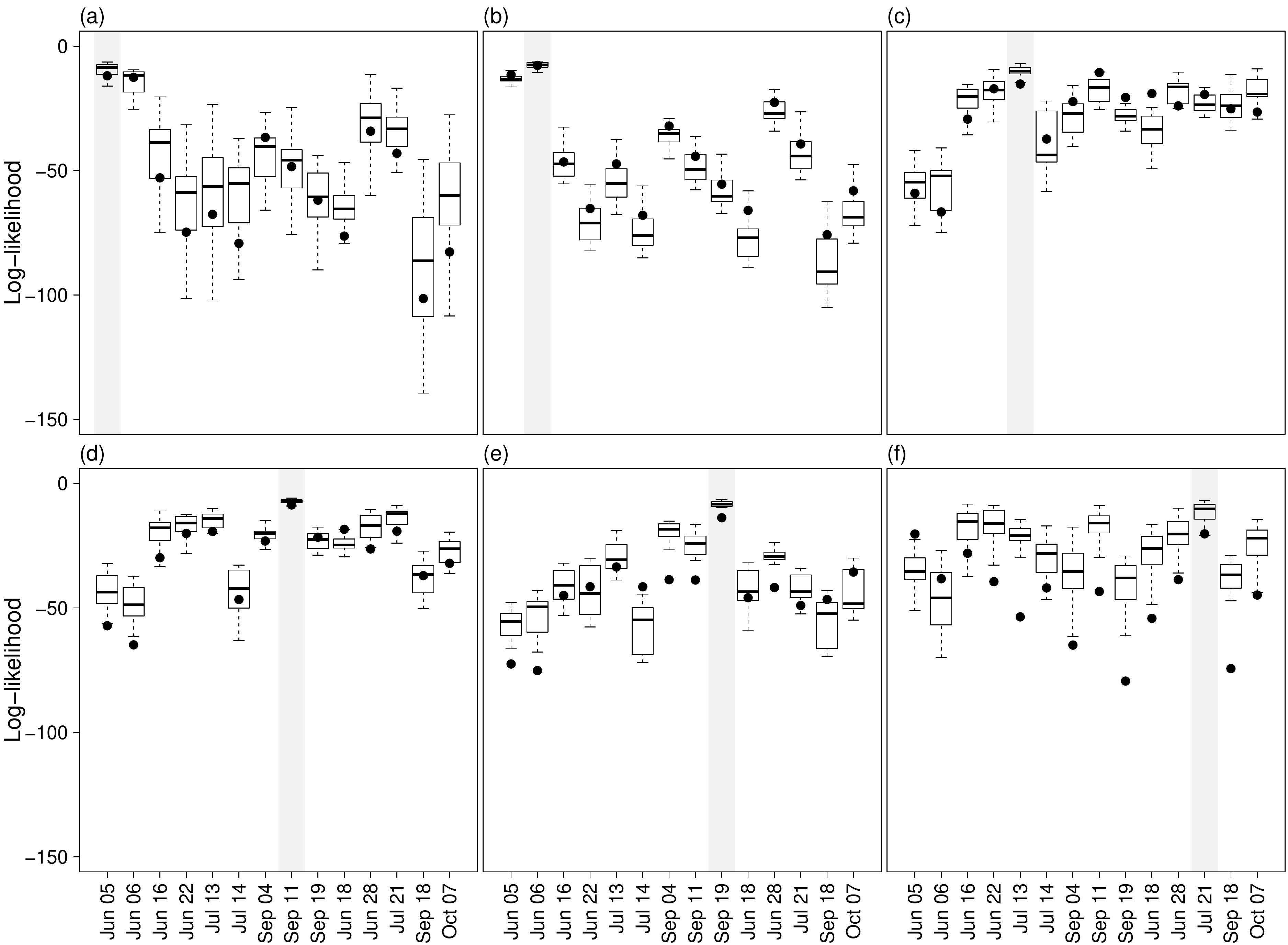}
 \caption{Log-likelihoods of the ensemble members $\log p_b( \mathbf m^{(j)}_k|C_i)$  over a 50 member cross-validation sample (box-whiskers) and the reanalysis $\log p_b( \mathbf o_i|C_i) $ averaged over the 50 member cross-validation sample (black dots) for member from class $k$ (x-axis) given class $C_i$. The class $C_i$ (gray shading) is (a) 5 June 2011, (b) 6 June 2011, (c) 13 July 2011, (d) 11 September 2011, (e) 19 September 2011, and (f) 21 July 2011.}
\label{Fig:loglikelihoodLDA4}
\end{figure}

{Fig. \ref{Fig:loglikelihoodLDA4} shows the log-likelihoods $\log p_b( \mathbf m^{(j)}_k|C_i)$ for a given class $C_i$, as well as  $\log p_b( \mathbf o_k|C_i)$ for the reanalysis. First of all{, as expected}, the log-likelihoods are generally maximal for $k = i$, which is indicated by the gray shading. For 5 and 6 June 2011, the log-likelihoods  are large for ensemble members from these days, with slight preference for the correct class, whereas all other ensemble members have significantly smaller log-likelihoods. The same holds for the log-likelihoods of the reanalysis. Conditionally on the other classes (Figs. \ref{Fig:loglikelihoodLDA4} (c)-(f)), the log-likelihoods of 5 and 6 June 2011 are particularly small, which indicates, that the two days have a significantly different spectral character (compare Tab. \ref{casestab}). Except for 21 July 2011, the log-likelihoods of the reanalyses are within the range of the  log-likelihoods of the ensemble members. Conditionally on class 21 July 2011, however, the spread of the log-likelihoods is larger and the log-likelihoods of the reanalyses are significantly smaller than those of the ensemble members. Still, the reanalysis of 21 July 2011 rates best in terms of log-likelihood for the 21 July 2011 class.}

\begin{figure}
 \includegraphics[width=\textwidth]{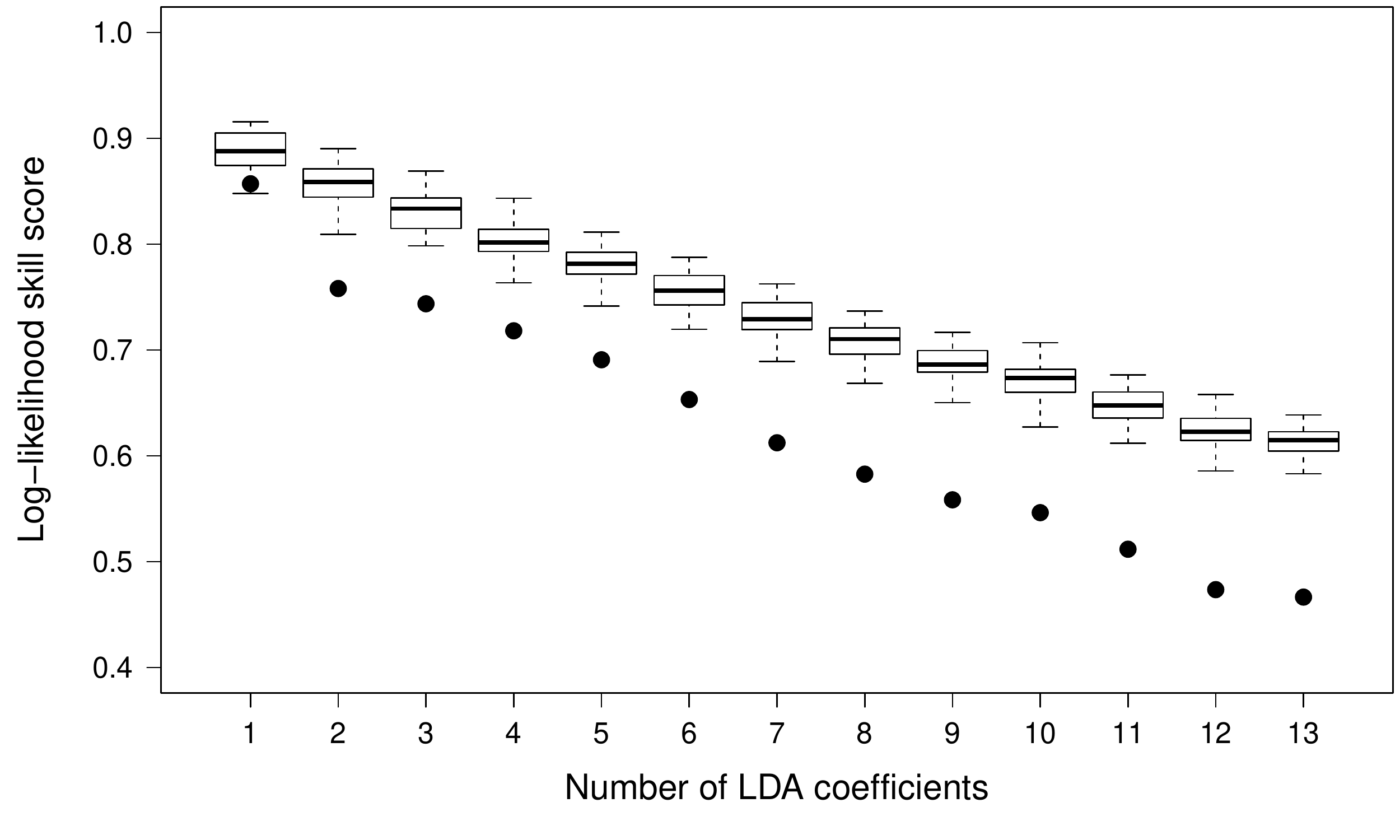}
 \caption{Log-likelihood skill scores with respect to the size of the LDA space. The box-whiskers represents the cross-validation range of the perfect forecasts skill $S_{perf}$, and the dots represent the averaged skill of the observations $S_{obs}$ both with respect to the reference score $S_{ref}$.}
\label{Fig:LSS} 
\end{figure}

{In a next step we investigate the skill score of the forecasts using
as reference $S_{ref}$ (Eq.~\ref{Eq:Sref}). Fig.~\ref{Fig:LSS} compares the range of skill estimates of the perfect forecasts $1- S_{perf,b}/S_{ref}$  and of the reanalysis $1-  S_{obs}/S_{ref}$ \ for a hierarchy of LDA spaces. First of all, the skill of the perfect forecasts linearly decreases with the number of LDA vectors, ranging from about 90\% using the first LDA vector only, to about 60\% using all 13 LDA vectors. This evolution may be understood by weighting the scores with the number of the LDA vectors. $S_{obs}/(\#LDA)$ is almost constant, whereas $S_{ref}/(\#LDA)$ increases with the number of the LDA vectors (not shown). While $p_b( \mathbf m^{(j_b)}_k|C_i), i\neq k$ is a normal distribution with covariance matrix $S_W + S_B$,  the covariance matrix of $p_b( \mathbf m^{(j_b)}_k|C_i), i= k$ is $S_W$. Thus $\det(S_W)$ linearly increases with the number of DOF, while $S_B$ much stronger than linearly decreases with the number of DOF. This is the effect of the LDA decomposition on $S_B$.
}

{
The skill of the reanalysis is within the uncertainty range of the perfect forecasts if only the first LDA vectors are used. It rapidly drops if 2 LDA vectors are used, which is partly due to the increasing difference for the reanalysis on 21 July 2011 (compare Fig.~\ref{Fig:loglikelihoodLDA4}). The skill of the reanalysis drops faster with the number of LDA vectors than expected from the perfect forecasts. It goes down to about 47\% for all 13 LDA vectors. This indicates, that the ensemble spread of the wavelet spectra does not contain the reanalysis at all scales and directions.
}

\subsection{Posterior assessment}

{Another way to assess the quality of the ensemble is to investigate how successfully new data $\mathbf x$ are attributed to the correct class, respectively day. As stated above, attribution uses the posterior probability $p_b(C_i|\mathbf x)$. LDA defines the subspace that maximizes discrimination within the ensemble, and per construction decreases the spread of the ensemble. This in turn results in a higher risk of a wrong attribution. In order to investigate the effect of overfitting, we estimate the attribution probability using again cross-validation (see section \ref{Sec:Method}\ref{Sec:MethodsLDA}).}

\begin{figure}
 \centering\includegraphics[width=1\textwidth]{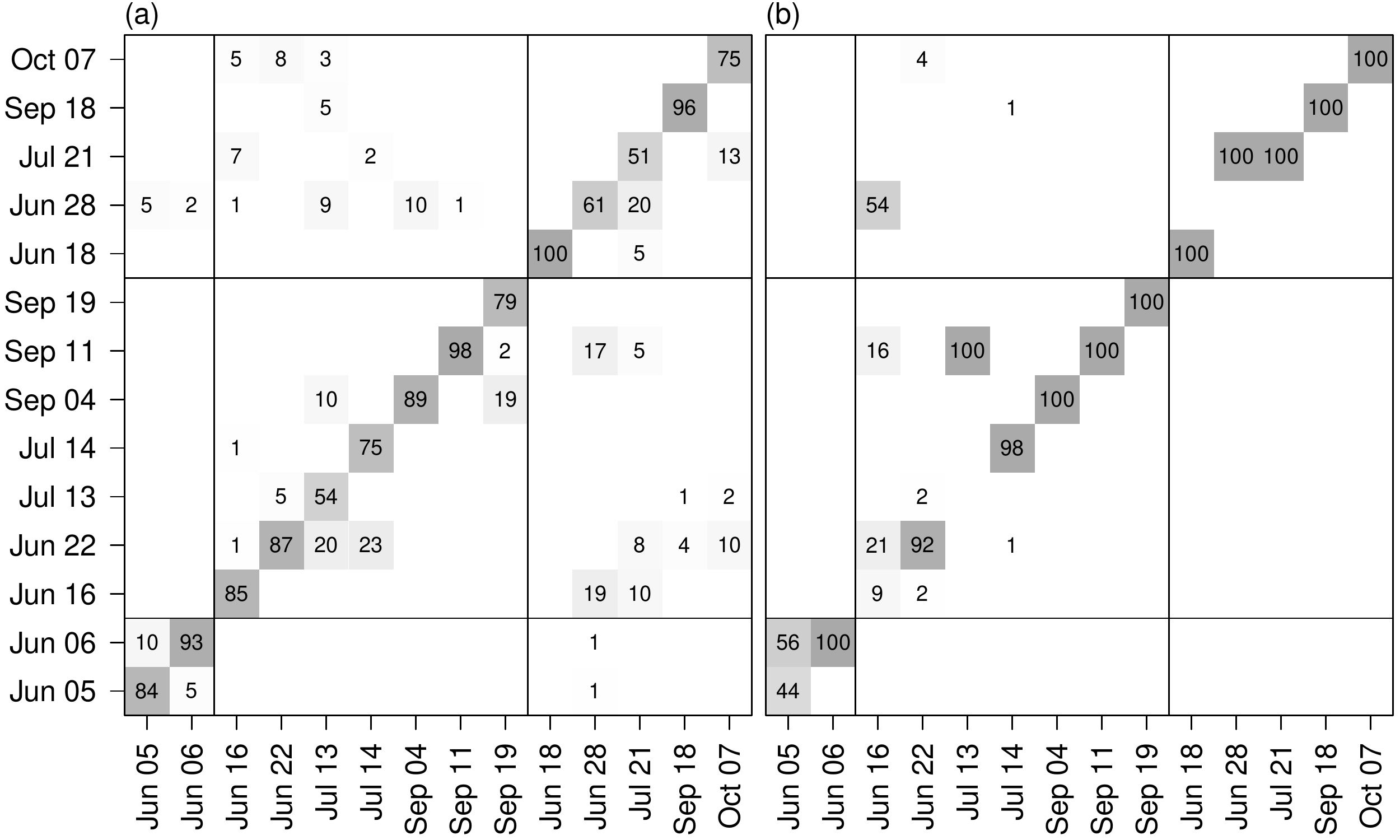}
 \caption{(a) LDA derived posterior probabilities of class $C_i$ (y-axis) given an ensemble member $\mathbf m^{(j_b)}_k$ (x-axis), namely $p_b(C_i| \mathbf m^{(j_b)}_k)$  in $\%$ using cross-validation. (b) LDA derived posterior probabilities of class $C_i$  (y-axis) given the reanalysis $\mathbf o_k$ (x-axis), $p_b(C_i| \mathbf o_k)$. For cross-validation one full day was excluded in the LDA estimation {and all 12 LDA coefficients were used.} Vertical and horizontal lines separate the different precipitation categories (C, F and FC).}
\label{Fig:LDAall}
\end{figure}

{Fig. \ref{Fig:LDAall} (a) shows the posterior probability of $p_b(C_i|\mathbf m^{(j_b)}_k)$ averaged over the cross-validation sample.} 
{A diagonal of $100\%$ indicates perfect discrimination.
The ensemble members show a posterior probability for the correct class of 80\% on average. For the two convective days, most probability is given to the correct class, and a probability of 5\% to 10\% for the respective other day. The mixed classes (FC) get some probability with respect to a few members of all classes, whereas none of the F classes would be attributed to the C members and vice versa.}

{Fig. \ref{Fig:LDAall} {(b)} shows the posterior probability of the reanalysis $p_b(C_i|\mathbf o_k)$. The posterior probability conditional on the reanalysis is sharper than for the ensemble members. The posterior probability prefers the correct class for most days. However, for 28 June 2011 and 13 July 2011 the posterior probability gives 100\% probability to the wrong class. 
The two convective days are very well attributed, although some probability is given to the other day, respectively. The reanalyses of 16 June 2011, 22 June 2011 and 14 July 2011 are not perfectly attributed to the correct class.}

\begin{figure}
 \centering\includegraphics[width=1\textwidth]{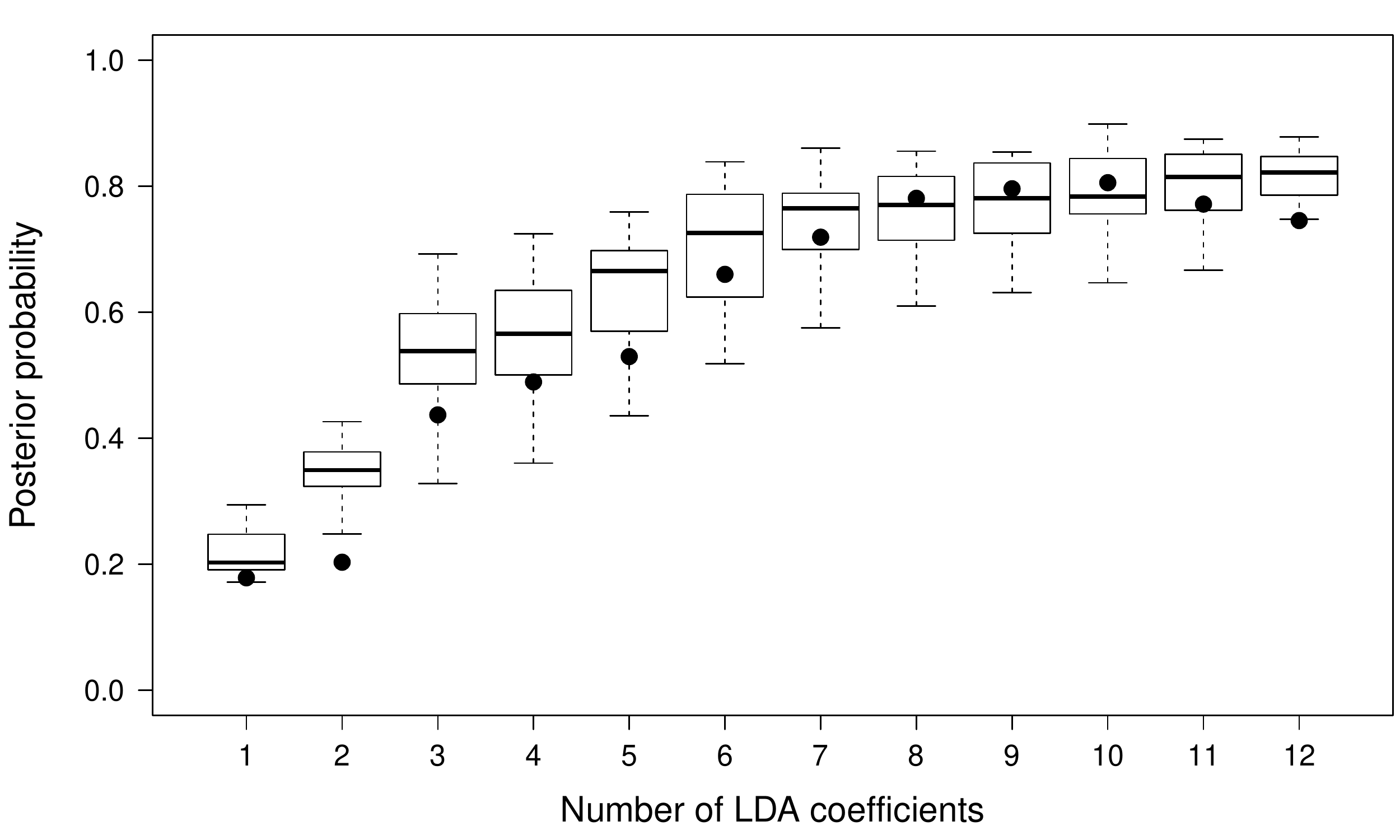}
 \caption{Averaged LDA derived posterior probabilities given 20 ensemble members $\sum_i p_b(C_i| \mathbf m^{(j_b)}_i)/N_c$, respectively (box-whisker over 20 members) and given the reanalysis $\sum_i  p_b(C_i| \mathbf o_i)/N_c$ (dots) using cross-validation as function of LDA vectors.}
\label{Fig:PostCV}
\end{figure}

{Fig.~\ref{Fig:PostCV} shows the average of the cross-validated posterior probabilities for the correct class as a function of the number of LDA vectors. With increasing number of LDA vectors, more details of the averaged wavelet spectrum are represented, and hence attribution probability increases with the number of LDA vectors. While with only the first LDA vector, merely 20\% of the  new data would be correctly attributed, the percentage increases to a maximum of about 80\% with 13 LDA vectors.
The averaged posterior probability of the reanalysis generally lies within the uncertainty range of the ensemble member. A maximum of above 80\% is reached with 10 LDA vectors. Thus although the log-likelihoods of the reanalysis are significantly smaller than those of the ensemble members, the reanalysis performs as good as the ensemble in terms of attribution.
}

\begin{figure}
 \includegraphics[width=\textwidth]
 {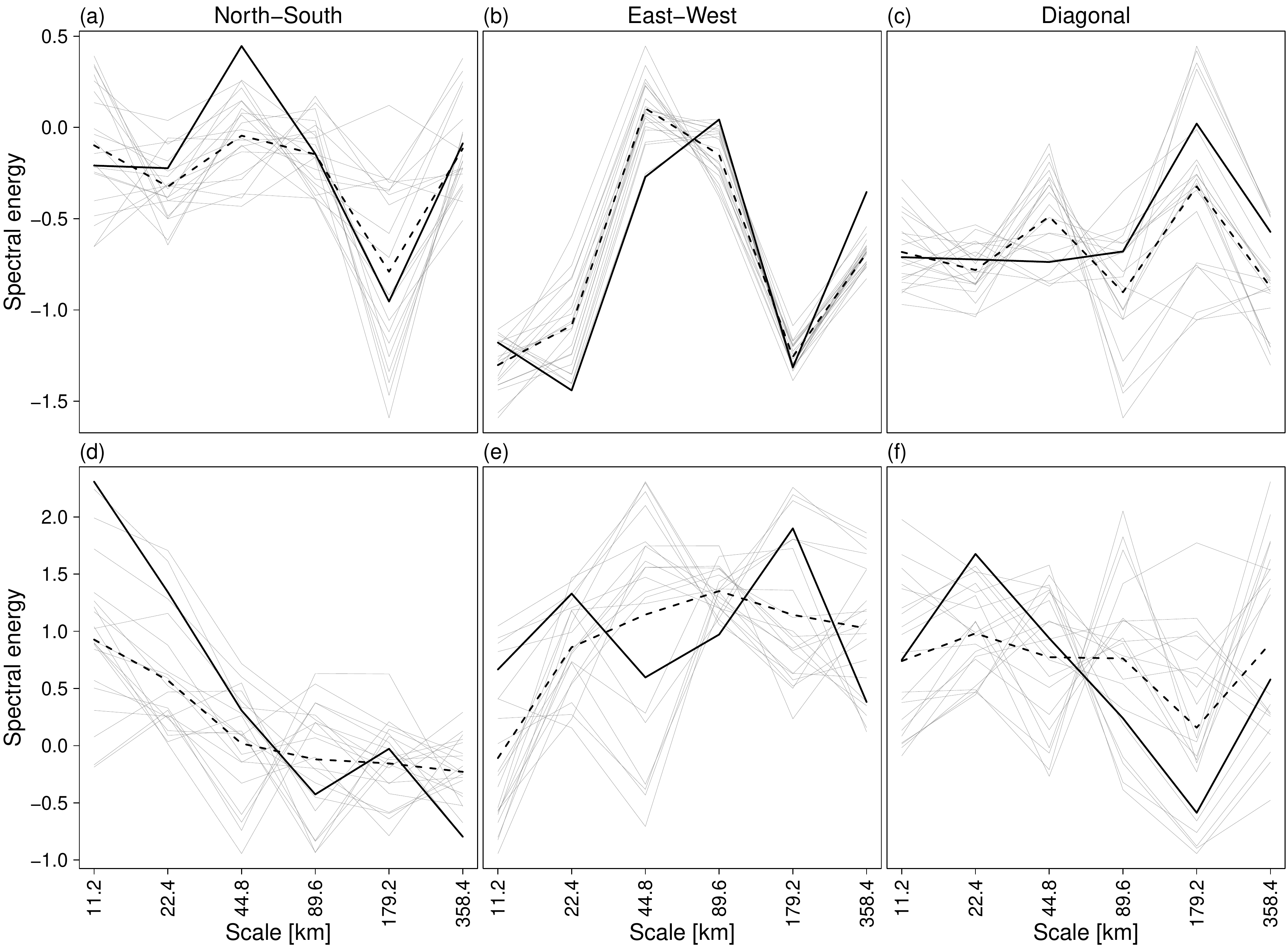}
 \caption{Filtered wavelet spectra for (a)-(c) 11 September 2011, and  (d)-(f) 21 July 2011 using 10 LDA vectors for the ensemble members (gray lines), and the reanalysis (solid black line). The dashed line indicates the class average over the filtered wavelet spectra of the ensemble members.}
\label{Fig:Spectrafiltered}
\end{figure}

{We finally have a look at the filtered wavelet spectra for two days. For the reanalysis of 11 September 2011 the likelihood as well as the posterior probability is large, which indicates a large resemblance between ensemble and reanalysis in terms of the averaged wavelet spectra. For 21 July 2011 the likelihood of the reanalysis is particularly small compared to the ensemble members, whereas the posterior probability is large.
Fig.~\ref{Fig:Spectrafiltered} shows the standardized wavelet spectra filtered using 10 LDA vectors. 
For both days, the filtered wavelet spectrum of the reanalysis lies within the range of the ensemble. The spread of the ensemble, however, largely varies. It is particularly small for the East-West direction on 11 September 2011, and large for 21 July 2011 (Note the different y-axes for the two days.). 
We remind the reader, that all classes share the same inner-class covariance matrix. Thus although the reanalysis on 21 July 2011 is consistent with the ensemble, the distance between the ensemble mean (class average) and the reanalysis is large with respect to the inner-class variance particularly for the small scales on North-South direction.
}

\section{Conclusions and discussion}\label{Sec:Conclusions}

{This study applies the technique for image texture analysis of \citet{eckley2010} to high-resolution precipitation ensemble forecasts. The averaged wavelet spectra are used to assess structural differences between ensemble forecasts and observations (i.e.~in our case reanalysis).
Further, data reduction is guided by a linear discriminant analysis that optimizes discrimination ability of the ensemble.
We use the subspace spanned by the LDA vectors to provide probabilistic forecasts of the averaged wavelet spectrum, and assess the skill score of a perfect forecast using for verification a forecast ensemble member against a non-perfect forecast using the reanalysis data.
The score differences point to deficiencies in the model where the reanalysis lies outside the range of the ensemble -- all with respect to the averaged wavelet spectra in a subspace spanned by the LDA vectors. {The causes of the deficiencies are not being exposed in this methodology, and the interpretation of the filtered wavelet spectra with respect to feature in the precipitation fields is not straightforward and would require additional investigation.}} 

{We further assess the discriminating power of the ensemble based on the averaged wavelet spectra. Correct attribution of new data, either from the ensemble or from the reanalysis is successful on average on 80\% of the data. 
The discriminating power for the reanalysis solely on the basis of the averaged wavelet spectra is remarkable, as it is almost as high as for the ensemble itself. This indicates a quite good resemblance between forecasts and reanalysis. Maybe the results are less remarkable when we take into account, that both forecasts and observations rely on a similar model, and that the forecast horizon is less than one day. But, on the other hand, it proves that the COSMO-DE-EPS model is a quite good forecast ensemble in this respect.
} 

{This study leaves room for further assessment and improvement. The data reduction using LDA is promising, as it favors data reduction in the directions, where the forecast ensemble has least discriminating power. However, this might impede the comparison of different forecasts ensemble, as they might have different directions in which they are most powerful in discriminating.
Improvements in the ensemble forecasts maybe obtained by using a variable inner-class covariance matrix, such that the ensemble spread is part of the probabilistic forecast. This would, however, require another reduction of DOF of the predictors, so that the estimated covariance matrix will not become singular.
Another issue is that the physical interpretation using an LDA filtered wavelet spectrum is not straight forward, and differences are sometimes hardly visible in the precipitation fields.}

{Finally, the true benefit of wavelet analysis is the localization of the wavelet spectrum. This has not been explored in this study. A local wavelet spectrum faces the same double penalty problems as other high-resolution fields, and deeper thoughts are needed to develop a multivariate verification approach on the basis of the local wavelet spectra.}

%%%%%%%%%%%%%%%%%%%%%%%%%%%%%%%%%%%%%%%%%%%%%%%%%%%%%%%%%%%%%%%%%%%%%
% ACKNOWLEDGMENTS
%%%%%%%%%%%%%%%%%%%%%%%%%%%%%%%%%%%%%%%%%%%%%%%%%%%%%%%%%%%%%%%%%%%%%
\begin{acknowledgment} 
Deutscher Wetterdienst, Offenbach, kindly provided the COSMO-DE-EPS forecasts.  Special thanks go to Sabrina Wahl for providing the  COSMO-REA2 reanalysis data which are produced within the Hans-Ertel Centre for Weather Research, University of Bonn. 
We highly appreciated the valuable comments of Barbara Casati and an anonymous reviewer, which helped us to significantly improve the manuscript.
We gratefully acknowledge financial funding by the project High Definition Clouds and Precipitation for Advancing Climate Prediction HD(CP)$^2$, funded by the German Ministry for Education and Research (BMBF) under the grants FKZ01LK1221A (Michael Weniger) and FKZ01LK1507B (Sebastian Brune).
\end{acknowledgment}

%%%%%%%%%%%%%%%%%%%%%%%%%%%%%%%%%%%%%%%%%%%%%%%%%%%%%%%%%%%%%%%%%%%%%
% APPENDICES
%%%%%%%%%%%%%%%%%%%%%%%%%%%%%%%%%%%%%%%%%%%%%%%%%%%%%%%%%%%%%%%%%%%%%
\ifthenelse{\boolean{dc}}
{}
{\clearpage}
\begin{appendix}[A]\label{App:Method}
We keep the general description of wavelets and their two-dimensional discrete version short. For a comprehensive introduction to wavelet transforms the reader is referred to \citet{daubechies1992}: ``\emph{Ten lectures on wavelets}''. For those new to the theory of wavelets we suggest \emph{Wavelets for kids} by \citet{vidakovic1994}, who describe the general context using few mathematically detailed formulations. Other introductions to wavelets are provided by \citet{vidakovic1999} and \citet{nason2008}.

\subsection{Two-dimensional non-decimated discrete wavelets}
\label{App:2DW}
A wavelet in one dimension is a function $\psi$ with compact support, that can be scaled in size and translated along the x-axis with a scale parameter $s \in \mathbb{R}^{+}$ and translation parameter $l \in \mathbb{R}$ by
\begin{equation}
\label{A1}
\psi_{s, l} \left( x \right) = \frac{1}{\sqrt{s}} \psi \left( \frac{x - l}{s} \right).
\end{equation}

$\psi$ is called the \emph{mother} wavelet, whereas the functions $\psi_{s,l}$ are called \emph{daughter} wavelets. The continuous wavelet transform of a signal $f$ is  given by the convolution of the daughter wavelets  $\psi_{s,l}$ with the signal $f$.

To obtain a discrete wavelet transform (DWT), $s$ in (\ref{A1}) is replaced by $s^{-j}_{0}$, where $s_{0} > 1$ is a constant and $j \in \Z^+$. The translation $l$  in (\ref{A1}) then depends on the scale using $kl_{0}s^{-j}_{0}$ with $k \in \mathbb{Z}$.
With $s_{0} = 2$ and $l_{0} = 1$ we obtain dyadic partitions, where the wavelets' size doubles (or halves) on each scale $m$, and the discrete mother wavelet {is scaled and translated as}
 \begin{equation}\label{A2}
\psi_{j, k} \left( x \right) = 2^{j/2} \psi \left( 2^{j} x - k \right).
\end{equation}
In contrast to \citet{nason2000} and \citet{daubechies1992} the formulation uses a positive scale index $j \in \Z^+$ as in \citet{eckley2010}. 

To define the two-dimensional wavelets a set of orthonormal functions $\phi_{j, k}$ is introduced,  which share {the same} multiscale relation as the mother wavelets \citep[dilation equation,][]{eckley2010}.\\
 The function $\phi(x)$ is called the {father wavelet}.
In contrast to $\psi(x)$ which has zero mean with $\int \psi(x)dx = 0$, the father wavelet is normalized as 
$\int \phi(x)dx = 1$. $\phi(x)$ can be interpreted as a low pass filter, whereas the mother wavelet acts as a high pass filter \citep{vidakovic1994}.
Discrete mother and father wavelets  $\psi_{j,k}$ and $\phi_{j,k}$ are constructed using the dilation equation.
For details the reader is referred to \citet{nason2000} and references therein. 

 For the extension from one to two dimensions we define a two-dimensional shift vector $\mathbf k = (k_1,k_2)$ with $k_1, k_2 \in \Z$. The 
 two-dimensional discrete wavelets have three directions, horizontal $h$, vertical $v$ and diagonal $d$. The respective two-dimensional wavelets are defined as 
 \begin{eqnarray}\label{A3}
 \mbox{horizontal: }&& \psi^h_{j,\mathbf k} =\phi_{j, k_1} \psi_{j, k_2} \nonumber\\
 \mbox{vertical: } && \psi^v_{j,\mathbf k} = \psi_{j, k_1} \phi_{j, k_2} \nonumber\\
 \mbox{diagonal:  } && \psi^d_{j,\mathbf k} = \psi_{j, k_1} \psi_{j, k_2},
 \end{eqnarray}  
 with $k_1, k_2 = 0, \ldots, L_j-1$, where $ L_j = (2^j-1)(N_h-1)+1$, where $N_h$ depends on the dilation equation of the wavelet family.
{If we were to sample only in steps of $2^j$, we would obtain an orthogonal wavelet basis as in Mallat’s multiresolution analysis \citep{mallat1989}. By translating to all locations at all scales instead, we obtain a redundant representation, which has the necessary robustness against shifts and noise in the signal.} Consequently, the set  of non-decimated two-dimensional wavelets is given as $\{ \psi^l_{j,\mathbf u}(\mathbf r) = \psi^l_{j,\mathbf u-\mathbf r} | \mathbf u,\mathbf r \in \Z^2 \}$.  
  The decomposition results in {a field of the same dimension as the input field but now} for each of the three directions horizontal ($h$), vertical ($v$), and diagonal ($d$), respectively \citep{fowler2005}. 

Autocorrelation wavelets of two-dimensional discrete wavelets are defined as the autocorrelation at lag $\bm \tau = (\tau_1,\tau_2)$ given as 
\begin{equation}\label{A4}
	\Psi_j^l(\bm\tau) = \sum_{\mathbf u \in \Z^2} \psi^l_{j,\mathbf u} (\mathbf 0) \psi^l_{j,\mathbf u} (\bm \tau). 
\end{equation}
The autocorrelation wavelets are separable and can be written in terms of the one-dimensional autocorrelations wavelets \citep[see][for a definition]{nason2000} analog to (\ref{A3}) as
 \begin{eqnarray}\label{A5}
 \mbox{horizontal: }&& \Psi_j^h(\bm\tau) = \Phi_j(\tau_1) \Psi_j(\tau_2) \nonumber\\
 \mbox{vertical: } && \Psi_j^v(\bm\tau) = \Psi_j(\tau_1) \Phi_j(\tau_2) \nonumber\\
 \mbox{diagonal:  } && \Psi_j^d(\bm\tau) = \Psi_j(\tau_1) \Psi_j(\tau_2).
 \end{eqnarray}

\subsection{The LS2W stochastic wavelet process and the local wavelet spectrum}

{The approach of \citet{eckley2010} provides estimates of the local covariance spectrum and uses them for a structure analysis. The approach is an extension of the one-dimensional analysis of \cite{nason2000} into two dimensions. } To this end, they define a stochastic locally stationary two-dimensional wavelet (LS2W) process 
 $X_{\mathbf R}(\mathbf r)$ as
\begin{equation}\label{A6}
	X_{\mathbf R}(\mathbf r) = \sum_{l\in \{h,v,d \}} \sum_{j=1}^\infty \sum_{\mathbf u \in \Z^2} w_{j,\mathbf u;\mathbf R}^l \psi_{j,\mathbf u}^l(\mathbf r) \xi_{j,\mathbf u}^l,
\end{equation}
where $\xi_{j,\mathbf u}^l$ is a zero-mean Gaussian variable with $\cov(\xi_{j,\mathbf u}^l , \xi_{j',\mathbf u'}^{l'}) = \delta_{j,j'}\delta_{\mathbf u,\mathbf u'} \delta_{l,l'}$, $w_{j,\mathbf u;\mathbf R}^l$ are coefficients for each location $\mathbf u$, direction $l$, and  scale $j$, and $\psi_{j,\mathbf u}^l(\mathbf r)$ are non-decimated wavelets on the two-dimensional discrete space $\mathbf r \in \{0, \ldots, R-1\}\times \{0, \ldots, S-1\}$. The spatial domain is defined by $\mathbf R = (R,S)$ with $R=2^m$ and $S=2^n$ and the smallest common scale $J(R,S) = \log_2(\min(R,S))$. {A smoothness assumption on the local amplitudes $w_{j,\mathbf u;\mathbf R}^l$ controls the local stationarity of the process. For details see \citep{eckley2010}.}

Now let  $\mathbf z = (z_1,z_2) \in (0,1)^2$ be the rescaled location such that $z_1 = r_1/R$ and $z_2 = r_2/S$.   Rescaling results in an increasing amount of information around a local structure as $\min(R,S) \to \infty$. This has the consequence, that under the assumption that the coefficients $w_{j,\mathbf u;\mathbf R}^l$ vary slowly {in space} (i.e. with $k$) the local structure becomes more and more stationary \citep[{\sl Remark 1} in][]{nason2000}. {The reasoning behind this is that once stationarity is reached, the local wavelet spectrum may be defined as the wavelet decomposition of the local autocovariance function.}

{Hence, \citet{eckley2010} define }the local wavelet spectrum (LWS) of the LS2W process {in (\ref{A6})}  as
\begin{equation}\label{A7}
	S_j^l(\mathbf z) = \lim_{\min(R,S) \to \infty} | w_{j,([\mathbf z \mathbf u]);\mathbf R}^l|^2,
\end{equation}
where $[\mathbf z \mathbf u]$ denotes the component wise product.
They show that the local autocovariance function $c(\mathbf z, \bm \tau)$ of an LS2W process with LWS $S_j^l(\mathbf z)$ is given as
\begin{equation}\label{A8}
	c(\mathbf z, \bm \tau) = \sum_{j=1}^\infty \sum_l S_j^l(\mathbf z) \Psi_j^l(\bm\tau),
\end{equation}
where $\Psi_j^l(\bm\tau)$ is the autocorrelation  wavelet defined in (\ref{A4}).
It can further be shown that the covariance of the LS2W process in (\ref{A6}) with $c_{\mathbf R}(\mathbf z, \bm \tau) = \cov(X_{ \mathbf R}([\mathbf z \mathbf R])X_{\mathbf R}([\mathbf z \mathbf R]+\bm\tau ))$ converges to $c(\mathbf z, \bm \tau)$ as $\min(R,S) \to \infty$. 
\citet{eckley2010} discuss the uniqueness of the formulation in (\ref{A4}) and show, that the inverse formula is
\begin{equation}\label{A9}
	S_j^l(\mathbf z) = \lim_{\min(R,S) \to \infty}\sum_i \sum_m (A^{l,m}_{j,i})^{-1} \sum_{\bm \tau} c(\mathbf z, \bm \tau) \Psi_i^m(\bm\tau),
\end{equation}
where $A^{l,m}_{i,j} =  \sum_{\bm \tau} \Psi_i^l(\bm\tau)\Psi_j^m(\bm\tau)$ define an operator $\mathbf A = (A^{l,m}_{i,j})_{i,j,l,m}$ with $i,j,=1,\ldots, J = \log_2(\min(R,S))$ and $l,m \in \{h,v,d \}$.

\subsection{Estimation of the local wavelet spectrum}

Empirical wavelet coefficients of an LS2W process are given by the projection 
\begin{equation}\label{A10}
	d_{j,\mathbf u}^l = \sum_{\mathbf r} X(\mathbf r) \psi_{j,\mathbf u}^l(\mathbf r).
\end{equation}
The two-dimensional wavelet periodogram given as $I_{j,\mathbf r}^l = |d_{j,\mathbf u}^l|^2$ is a biased and inconsistent estimator of $S_j^l(\mathbf z)$. Thus smoothing is needed to obtain a consistent estimator. In \citet{eckley2010} and \citet{eckley2011} this is obtained by non-linear wavelet shrinkage. For details the reader is referred to \citet{eckley2010} and references therein.
The bias is a consequence of the redundancy of the non-decimated wavelet representation which introduces autocorrelations between the wavelets. An unbiased estimator is thus calculated on the smoothed periodogram $\tilde I_{j,\mathbf r}^l$ by weighting the periodogram with the inverse of the operator matrix $\mathbf A$. The reader is again referred to \citet{nason2000} and \citet{eckley2010} for details.

\end{appendix}

%%%%%%%%%%%%%%%%%%%%%%%%%%%%%%%%%%%%%%%%%%%%%%%%%%%%%%%%%%%%%%%%%%%%%
% REFERENCES
%%%%%%%%%%%%%%%%%%%%%%%%%%%%%%%%%%%%%%%%%%%%%%%%%%%%%%%%%%%%%%%%%%%%%
% Create a bibliography directory and place your .bib file there.
\ifthenelse{\boolean{dc}}
{}
{\clearpage}
\bibliographystyle{ametsoc}
\bibliography{bibliography.bib}

\begin{table}[ht]
\caption{Selected forecast dates in 2011 with time of day and precipitation type (C = convective, F = frontal, FC = mixture of convective and frontal precipitation). All forecasts are initialized at 00 UTC.}
\centering
\label{casestab}
\begin{tabular}{llllp{10cm}}
  \hline
&Date & UTC & & Character \\
  \hline
A:&Jun 05 & 16 & C & Many scattered thunderstorms in Western Germany, triggered by surface wind convergence due to thermal lows. \\
B:&Jun 06 & 16 & C & A huge number of scattered cells over complete Germany. \\
C:&Jun 16 & 16 & F & North-South orientated cold front in Central Germany with single cells in front and behind. \\
D:&Jun 22 & 16 & F & A cold front covers most parts of Germany with embedded convection. \\
E:&Jul 13 & 16 & F & Weak frontal precipitation over the complete domain with a few convective cells within. \\
F:&Jul 14 & 18 & F & Synoptic scale rainfall in Benelux and single cells in Southeastern Germany. \\
G:&Sep 04 & 18 & F & A narrow North-South orientated front covers Central Europe from Denmark to Switzerland.  \\
H:&Sep 11 & 18 & F & Strong synoptic scale rain pattern in Central Germany with small cells in Western Germany. \\
I:&Sep 19 & 16 & F & Large scale precipitation in Austria and small cells in Denmark as well as in Switzerland. \\
J:&Jun 18& 16 & FC & Synoptic scale rainfall close to the Alps with scattered convection in Northwestern Germany. \\
K:&Jun 28 & 18 & FC & Convergence line with more stratiform rain in Northern part and convective pattern in Southern parts along Benelux and Eastern France. \\
L:&Jul 21 & 16 & FC & Front in Eastern Germany with scattered convection in Southwestern Germany. \\
M:&Sep 18& 16 & FC & Large scale rainfall along a cold front from Baltic Sea to the Alps. Some weak cells behind. \\
N:&Oct 07& 16 & FC & Many small scale convective cells in Germany with large scale precipitation in southeastern corner of the domain. \\
  \hline
\end{tabular}
\end{table}

\end{document}